\documentclass{article}

\usepackage{arxiv}

\usepackage[utf8]{inputenc} 
\usepackage[T1]{fontenc}    
\usepackage{hyperref}       
\usepackage{url}            
\usepackage{booktabs}       
\usepackage{amsfonts}       
\usepackage{nicefrac}       
\usepackage{microtype}      
\usepackage{lipsum}
\usepackage{graphicx}
\usepackage{enumitem}
\usepackage{amsmath}
\usepackage{algorithm}
\usepackage[noend]{algpseudocode}
\usepackage{multirow}
\usepackage{natbib}
\makeatletter
\def\BState{\State\hskip-\ALG@thistlm}
\makeatother

\title{Void Region Segmentation in Ball Grid Array Using U-Net Approach and Synthetic Data}

\author{
 Vijay Kumar Neeluru\\
  \texttt{vijaykumar413@gmail.com} \\
   \And
 Vikas Ahuja\\
  Intel Corporation (India) Pvt. Ltd., \\
  Bangalore, India\\
  \texttt{vikas.ahuja@intel.com} \\
}

\begin{document}
\maketitle

\footnotetext[1]{This work was done while the first author was working for Intel India, Bangalore.}

\begin{abstract}
The quality inspection of solder balls by detecting and measuring the void is important to improve the 
board yield issues in electronic circuits. In general, the inspection is carried out manually, based on 
2D or 3D X-ray images. For high quality inspection, it is difficult to detect and measure voids accurately 
with high repeatability through the manual inspection and the process is time consuming. In need of high quality 
and fast inspection, various approaches were proposed, but, due to the various challenges like 
vias, reflections from the plating or vias, inconsistent lighting, noise and void-like artifacts makes these 
approaches difficult to work in all these challenging conditions. In recent times, deep learning approaches are 
providing the outstanding accuracy in various computer vision tasks. Considering the need of high quality and 
fast inspection, in this paper, we applied U-Net to segment the void regions in soldering balls.
As it is difficult to get the annotated dataset covering all the variations of void, we proposed an approach to 
generated the synthetic dataset. The proposed approach is able to segment the voids and can be easily scaled to 
various electronic products.     
\end{abstract}


\section{Introduction}
In the electronic circuits manufacturing, a type of surface-mount named soldering ball grid array (BGA) is 
being used widely. Due to various reasons the voids will appear in these soldering balls that will reduce 
the life time of the device. Inspecting the quality of BGA to confirm the presence of voids is vital for quality 
inspection and to reduce the cost of manufacturing ~\cite{Hill2011}. In general, 2D and 3D X-ray imaging is used and 
voids are segmented based on the intensity differences within the BGA. Due to the low contrast between 
the voids and the BGA, 
it is difficult to accurately segment the void and an human expert is needed for the reliable inspection. The issue with
the human inspection is the low repeatability due to the perceptual differences. As the number of soldering balls in 
BGA ranges from 10s to 100s, 
each ball is to be inspected in serial fashion resulting in longer inspection time. Due to these challenges in manual 
inspection, there is a need for low delay robust void detection approach that can be scaled to all 2D X-ray acquisition 
device type, product type, layout of BGA and the void characteristics. 
  
Image processing techniques are usually applied for void detection by segmenting each soldering ball and 
segmenting the void regions within each soldering ball. The arrangement, size, intensity and number of soldering balls 
will vary for different 
devices. The challenging factors in void segmentation are relative intensity of voids, vias, plated-through holes, 
reflections from the plating or vias, inconsistent lighting, background traces, noise, void-like artifacts, and 
parallax effects. There is a need of robust techniques that are able to segment soldering balls and voids considering
all the mentioned factors.
Although many techniques are proposed since many years, but lacks in the robustness in dealing these various challenges.
The drawbacks of these approaches are scalability to new device types or different void 
characteristics and the parameters has to be tuned manually for different device and material types. 

In ~\cite{Said2012}, Laplacian of Gaussian (LoG) is applied to detect edges, via extraction, filtering out false voids 
and finally detect real voids. In ~\cite{Peng2012}, blob filters with various sizes are applied to segment the voids
of various sizes. In 
~\cite{Mouri2014}, void detection is formulated as the matrix decomposition problem by assuming that void as sparse 
component and non-negative matrix factorization approach is used to separate the void region from soldering ball. 

The end to end void inspection process can be broadly split into two steps. In the first step, the soldering balls are 
segmented from the input image and in the second step, the voids are to be segmented and analysed for each soldering 
ball. 

\textbf{Soldering Ball Segmentation}

In the first step, thresholding or background subtraction techniques are usually applied to remove the background, 
that works well to separate out the soldering balls from background. Some of the soldering balls are 
not detected due to shadowing of other components. In these cases, reference ball based matching technique is 
applied in ~\cite{Said2010}. Where, the reference template is used to match all the locations in the neighbourhood 
of occluded soldering ball. The best matched location is considered as the location of occluded one. In general, 
the pattern or arrangement of soldering balls will vary for each manufacturing device, prior information about the 
soldering balls pattern can simplify the soldering ball segmentation. To solve the pattern issue various approaches 
are proposed in ~\cite{Said2010}.   
 
\textbf{Void Detection} 
   
The challenging factors in void segmentation are relative intensity of voids, shape of void, vias, plated-through holes, 
reflections from the plating or vias, inconsistent lighting, background traces, noise, void-like artifacts, and 
parallax effects. The non-void soldering balls are of uniform intensity and voids are brighter with respect to 
intensity of non-void 
soldering ball. The shape of voids are assumed as circular in ~\cite{Said2010} to simplify the detection. 
But, in practical, the voids can be of regular and irregular shapes and the 
overlapping of multiple regular voids will look as irregular void. The issue with the irregular shapes is 
difficulty in distinguishing the irregular shape voids and overlapping voids. Normally, multiple voids will 
appear. In some cases, the voids are partially or totally obscured by vias and the characteristics 
of via reflections are similar to those of the actual voids. The challenges are the scalable robust void segmentation 
irrespective of shape, number of voids, overlapping between multiple voids, and noises. In recent years, deep learning 
techniques are 
showing the 
promising accuracy for various computer vision problems such as image classification, object detection and pixel level 
segmentation. In this paper, we apply deep learning approach by formulating the void detecting problem as segmentation 
problem. Some of the popular segmentation approaches are U-Net ~\cite{Ronn2015}, Mask R-CNN ~\cite{He2017}, 
FCN~\cite{Shel2017}, U-Net++~\cite{Xue2018} and Adversial 
U-Net ~\cite{Zhou2018}. U-net architecture consists of encoder and decoder network with skip connections and is widely 
used in various segmentation problems. Deep learning approaches for void segmentation are not explored. In this paper, 
we propose U-Net based approach for void regions segmentation. To detect the 
soldering ball we followed the same approach used in ~\cite{Said2012} with some modifications. For training the U-Net, 
an large annotated dataset is needed resembling all the variations in voids. As it is difficult to get the void dataset
covering all the variations. Recently, synthetic dataset generation became the vital thing. In this paper, we proposed 
the approach to generate the synthetic dataset to solve the data problem. According to our knowledge, this is the
first paper to propose the technique for generating the synthetic data for void detection. Initially we train the 
2-class binary 
image classification network (U-net encoder) to classify whether the image consists of void or not. U-net encoder 
weights are initialized with this and end to end U-net (Encoder + Decoder) is trained for void segmentation. 
The paper is organized as follow. The proposed approach is 
described in section 2, experimental results are presented in section 3 and finally concluded in section 4.

\section{Proposed Approach}
\label{sec:headings}

\begin{figure}[h]
\centering
\includegraphics[width=0.9\textwidth, height = 0.4\textheight]{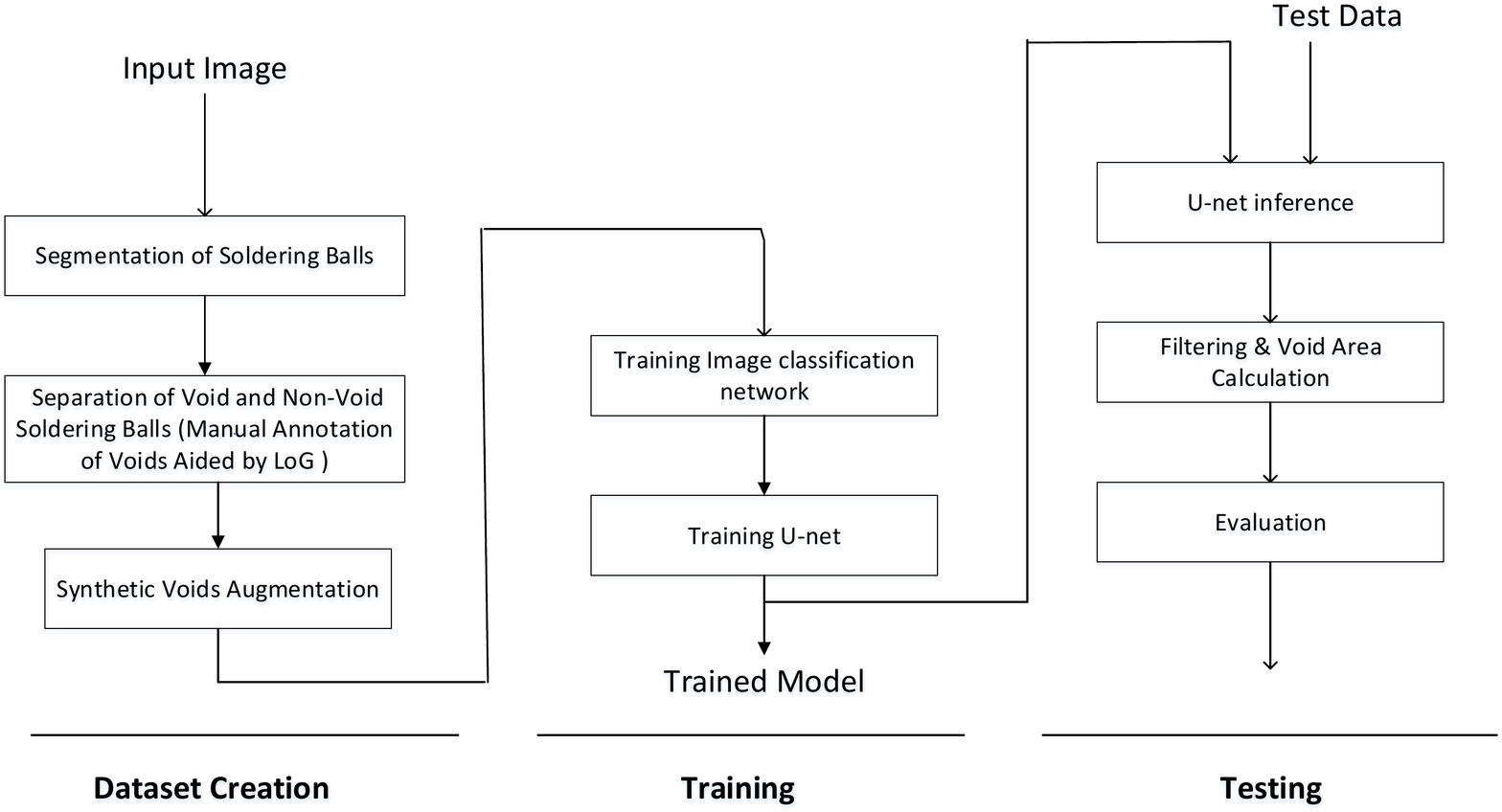}
\caption{The Flowchart of Proposed Approach Consisting of Dataset Creation, Training and Testing}
\label{fig:overall_flowchart}
\end{figure}

The proposed approach mainly consists of 3 stages as shown in the Fig. ~\ref{fig:overall_flowchart}. 
The first stage is the dataset creation. Given the input image,
soldering balls are to be segmented. For each soldering ball, the voids are to be
manually annotated aided by LoG, and classified into void or non-void types. Synthetic voids are augmented on the
non-void soldering balls to generate the dataset for training U-net. In the second stage, image classification
network (U-net encoder) is trained for void/non-void classification and end to end U-net (Encoder + Decoder) is 
trained for void segmentation. Finally, testing is applied on the test dataset for evaluation including with post 
processing step to remove the noises. The detailed description of these stages are 
described next. 

\subsection{Dataset Creation}

\subsubsection{Segmentation of Soldering Balls}
Given an input image, ~\cite{Said2012} applied adaptive thresholding, circle detection and interpolating the occluded 
balls. The template matching approach is used to segment the occluding balls. The reference soldering ball is used to 
match all the locations in the occluded region. The location having the higher correlation is considered as the centre 
of the occluded soldering ball. In this paper, we follow the similar steps with some modifications. The steps in our 
proposed approach consists of slicing, otsu thresholding, circle fitting, filtering the false voids and interpolating 
the occluded balls. In this paper, we proposed the location based filtering using the information of un-occluded 
soldering balls. If the ball is missing in detection step then the distance between the neighbouring balls will be 
more. We interpolate the location of occluded one based on the distances of neighbouring soldering balls. 
The pseudo code of proposed algorithm is given in Algorithm \ref{Sol_seg}. 

\begin{algorithm}
\caption{Soldering Ball Extraction}\label{Sol_seg}
\begin{algorithmic}[1]
\State Given an input image, divide it into non overlapping slices of size 300x400
\State For each slice, apply otsu thresholding
\State Rearrange the segmented slices in the corresponding position in the image to get the segmented image
\State Given the segmented image, find the circles and for each circle get the circle location and radius
\State Remove the circles (non-soldering balls) and interpolate the missing ones 
\State Find the mode of the radius's ($r_{mode}$), remove the circles (non-soldering balls) whose absolute 
		difference between the $r_{mode}$ and the radius ($r$) is greater than $r_{thr}$. |$(r_{mode}-r)$|>$r_{thr}$ 
		$r_{thr}$ = $r_{mode}$/sca. sca = 5 is selected heuristically that gives the best results. 
\State Cluster the soldering balls based on the horizontal and vertical position. Each cluster will represent the
group of soldering balls which are aligned in horizontal and vertical position 
\State In case of ideal detection, the number of soldering balls in each cluster will be equal in horizontally and 
vertically. The horizontal and vertical between the soldering ball and neighbouring ones will be equal. The reference 
distance $d_{ref}$ is calculated for each soldering ball along horizontally and vertically.  
\State In each cluster, find the distances between the neighbouring soldering balls. In case of missing ones, the 
distance (d) between the neighbouring ones will be more compared to the $d_{ref}$. 
\State The location ($c_{x}$,$c_{y}$) of missing ones are interpolated using the neighbouring ones and using the 
$d_{ref}$. 
\State After finding the ($c_{x}$,$c_{y}$), the reference (neighbour soldering balls) are used as template and a 
full search [-SR,SR]x[-SR,SR] is performed around the location ($c_{x}$,$c_{y}$). The location which gives the minimum 
error is considered as the optimal location. 
\end{algorithmic}
\end{algorithm}

\subsubsection{Separation of Void and Non-Void Soldering Balls}

Given the images of soldering balls, the void contours are generated through manual annotation.  
For this, LoG is applied at first to get the void contours. The LoG contours are 
classified into two categories, opened and closed. For closed contours, The region inside the contour represents the 
void and no manual labelling is required for these cases.  Whereas, for opened contours, the contours are to be 
closed by manually labelling the most probable pixels that forms the closed contour. 
To do that, the input images, LoG contours are manually analysed by profiling the intensity values along horizontal 
and vertical around the void using ImageJ to find the most probable pixels that forms the closed contour.  
In case of overlapping contours, the contours are to be separated into individual ones. 
The manual process required is to separate the individual contours in case of overlapping contours. 
Given the input image and LoG mask, the overall manual process is to find the most probable pixels that can form the 
closed contour in case of open contours and separate the individual contours in case of 
overlapping contours. The voids with various strength usually appear in soldering balls, but the voids whose intensity 
is more than certain threshold $Thr_{min}$ are considered as real voids. For each contour, the average intensity
of the pixels inside the contour is calculated representing the intensity of void ($I_{void}$). The average
intensity of the pixels surrounding the void are calculated representing the background intensity ($I_{BG}$). 
Voids satisfying the condition ($I_{void}$ - $I_{BG}$) < $Thr_{min}$ are considered as invalid voids and these contours
removed. After that, steps described in Algorithm \ref{GB} are applied to separate the soldering balls into 
void or non-void.


\begin{algorithm}
\caption{Separation of Void and Non-Void Soldering Balls}\label{GB}
\begin{algorithmic}[1]
\State Given the contour image, check for existence of any closed contour that indicates the presence of void
\If {\text{ any closed contour is found }}
\State  \text{then classify it as void soldering ball}
\Else  {\text{ Classify it as non-void soldering ball}}
\EndIf
\end{algorithmic}
\end{algorithm}

\subsubsection{Synthetic Voids Augmentation}

It is widely known that, large dataset is the key to solve the problem using deep learning. Millions of images were 
needed to solve the image classification problem and became the backbone network to solve the various vision 
problems ~\cite{HZRS2015}. Training the network using less number of samples can results in over-fitting that 
leads to poor accuracy. Recently, many novel approaches were proposed to solve the data and labelling problem 
for various complex computer vision tasks. As the millions of labelled samples may not be available in most of 
the cases, data augmentation, transfer learning ~\cite{Tan2018}, domain adaptation ~\cite{judy2017} and synthetic 
dataset creation became the vital approaches to improve the accuracy ~\cite{pere2017}. The simple photometric and 
geometric transformations like translation, rotation, scaling, noise etc., were used for 
data augmentation. Recently, GANs are more popular in generating synthetic samples that looks like real and are 
widely used in data augmentation, creation of adversarial examples and image translation. Unsupervised learning 
~\cite{anto2017} techniques are used for learning features, if the labelled data is not available. 

As it is difficult to get the void samples 
covering all the diversities (location, intensities and shape of voids, number of voids). Although data augmentation 
techniques are usually applied, but it easily leads to over-fitting and doesn't increase the sample diversity for void
segmentation. 
As from the manual annotation it is observed that number of non-void samples 
are more and the voids are of uniform intensity. There is a possibility to generate the synthetic voids with 
various intensities, sizes, count, locations and augment them on the non-void soldering balls that looks like 
real voids. This approach will create the more diverse samples and generalize the void detection that provides 
the better accuracy in real test time. The real void samples characteristics has to be considered in generating 
the synthetic voids so that the samples generated using synthetic domain should look as real. The characteristics 
to be considered in creating the synthetic voids are mentioned below.

Characteristics of the void to be considered in generated the dataset are
\begin{itemize}
\item Location of Voids: The voids can occur any location in soldering ball. The dataset should consists of voids 
in all possible locations
\item Size of Voids: Voids can occur at any location. The dataset should consists of void with possible various sizes
\item Number of Voids. The dataset should consists of with varying number of voids
\item Brightness of Voids: The brightness of void can be of any intensity, some are less brighter and some are more. 
	The dataset should consists of voids with various brightness
\item Filtering between void and background: The intensity transition between ideal synthetic soldering ball and 
background is ideal. In practical, the transition is gradual. The low pass filter has to applied
for the voids so that the edges of synthetic voids looks real. For generalization various low filters (amount of blur) 
has to be applied.    
\item The generated synthetic voids will be of uniform intensity with out any variations. To create, small intensity 
variations within the void, noise with different variances has to be added.  
\item Occurrences of multiple overlapping and non-overlapping voids are common, the dataset
	should consists of all possible combinations of multiple voids  
\end{itemize}

\begin{algorithm}
\caption{Generation of Synthetic Dataset for Training U-net}\label{Aug}
\begin{algorithmic}[1]
\State The following steps [2-14] are repeated for $I_{max}$ number of times
\State Sample one of the non-void soldering ball
\State Sample a random number void count (VC) from the range [$VC_{min}$ $VC_{max}$] representing the number of voids
\State Sample the  number of random values from the range [$VR_{min}$ $VR_{max}$] representing the void radius(VR) for 
each void
\State Sample the VC number of random values from the range [$VI_{min}$ $VI_{max}$] representing the void intensity(VI)
\State Sample the VC number of random values from the range [$VB_{min}$ $VB_{max}$] representing the void blur(VB)
\State Sample the VC number of random values from the range [$VN_{min}$ $VN_{max}$] representing the void noise 
 (VN) variance
\State Sample the VC number of random values from the range [0 H] representing the void x-locations (VX) for each void
\State Sample the VC number of random values from the range [0 W] representing the void y-locations (VY) for each void
\State Generate the VC number of circular voids with the corresponding intensity VI, radius VR. The 
circular generation step consists of copying the intensity VI in all the pixel locations where the distance between the centre 
and the pixel location is less than radius VR.    
\State In practical, voids will be appearing inside the soldering ball. In this approach, there is a possibility of 
voids crossing the boundary of soldering ball and is considered as invalid and shouldn't be augmented. If any of the 
void crosses the boundary of soldering ball then remove the corresponding void.
\State Generate gaussian noise using the sampled noise variances VN and add to the generated circular voids. 
\State Augmenting the voids to the soldering creates a sharp boundary between the void and background. In practical, 
the transition between the two void and background has to be smooth. So, we augment the each void on the sampled 
soldering ball blur the edges of voids with the sampled blur factors VB.
\State Add the void augmented soldering ball to the dataset

\end{algorithmic}
\end{algorithm}

\begin{figure}[h]
\centering
\includegraphics[width=0.6\textwidth, height = 0.4\textheight]{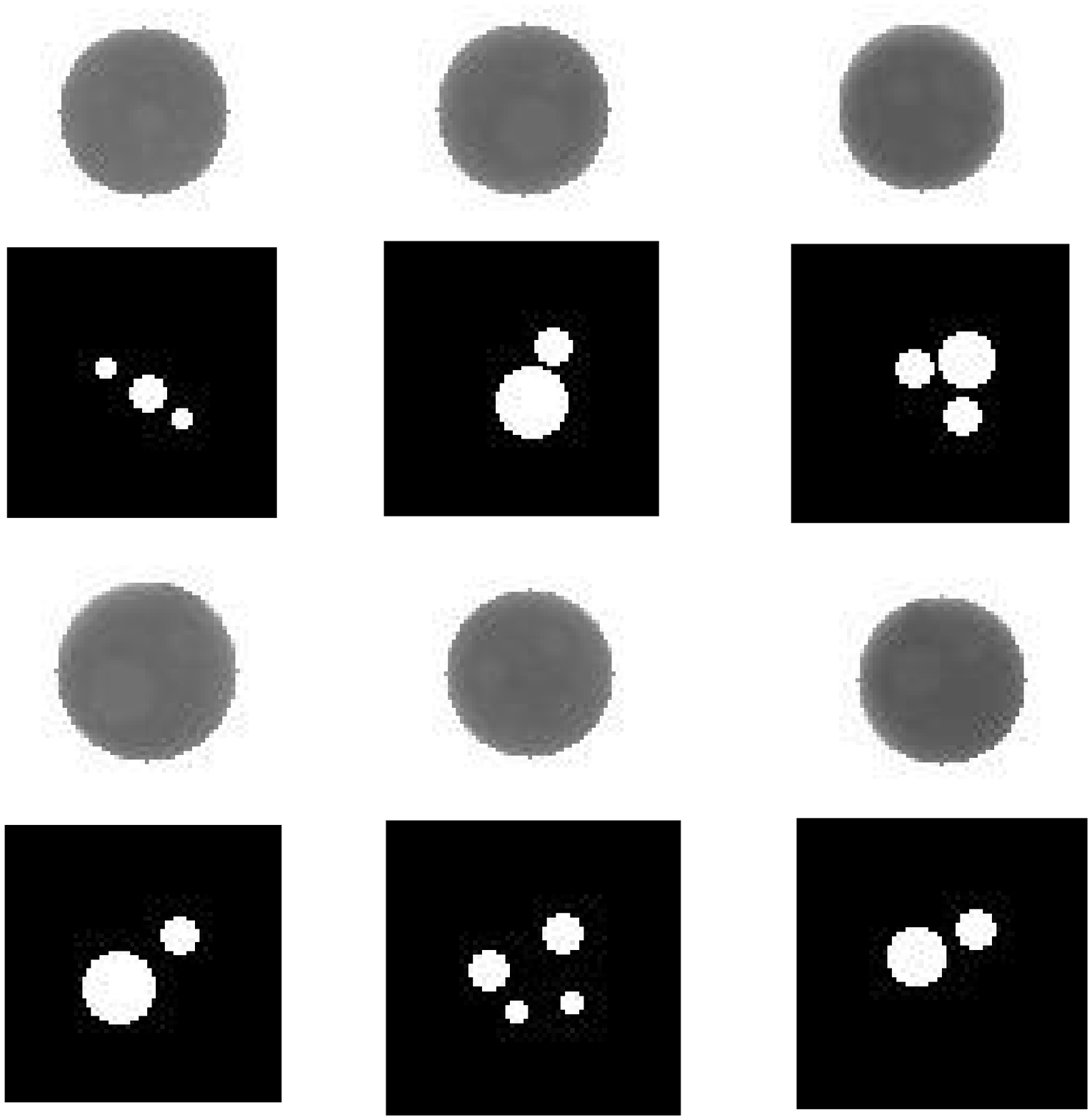}
\caption{Sample Images of Artificial Voids Created on the Non-Void Soldering Balls}
\label{fig:sample_aug_voids}
\end{figure}

In this paper we considered all these mentioned factors in creating the diverse synthetic dataset. Regarding the 
shape of void, we assume that voids are circular. The factors considered here are void location (VL), void radius (VR), 
void brightness (VB), amount of void blur (VBL), 
void noise (VN) variance and number of voids or void count (VC). 
Given the minimum and maximum ranges of these values, we randomly pick one value for each of them and generate the 
void and augment on the soldering ball. Given the allowable 
minimum ($VR_{min}$) and maximum void radius ($VR_{max}$), 
minimum ($VC_{min}$) and maximum void count ($VC_{min}$),
minimum ($VB_{max}$) and maximum ($VB_{max}$) void brightness, 
minimum ($VBL_{min}$) and maximum ($VBL_{max}$) blur factor, 
minimum ($VN_{min}$) and maximum ($VN_{max}$) noise variance, 
maximum images to be generated ($I_{max}$), 
image height (H) and width (W) and the images of non-void 
soldering balls, the procedure described in Algorithm \ref{Aug} is applied to generate the synthetic dataset. 
The sample 
images of void augmented soldering balls and the corresponding void masks are shown in Fig. \ref{fig:sample_aug_voids} 

\subsection{Training}

\subsubsection{Training Image Classification Network (Encoder)}
\begin{figure}[h]
\includegraphics[width=1.2\textwidth, height = 0.6\textheight]{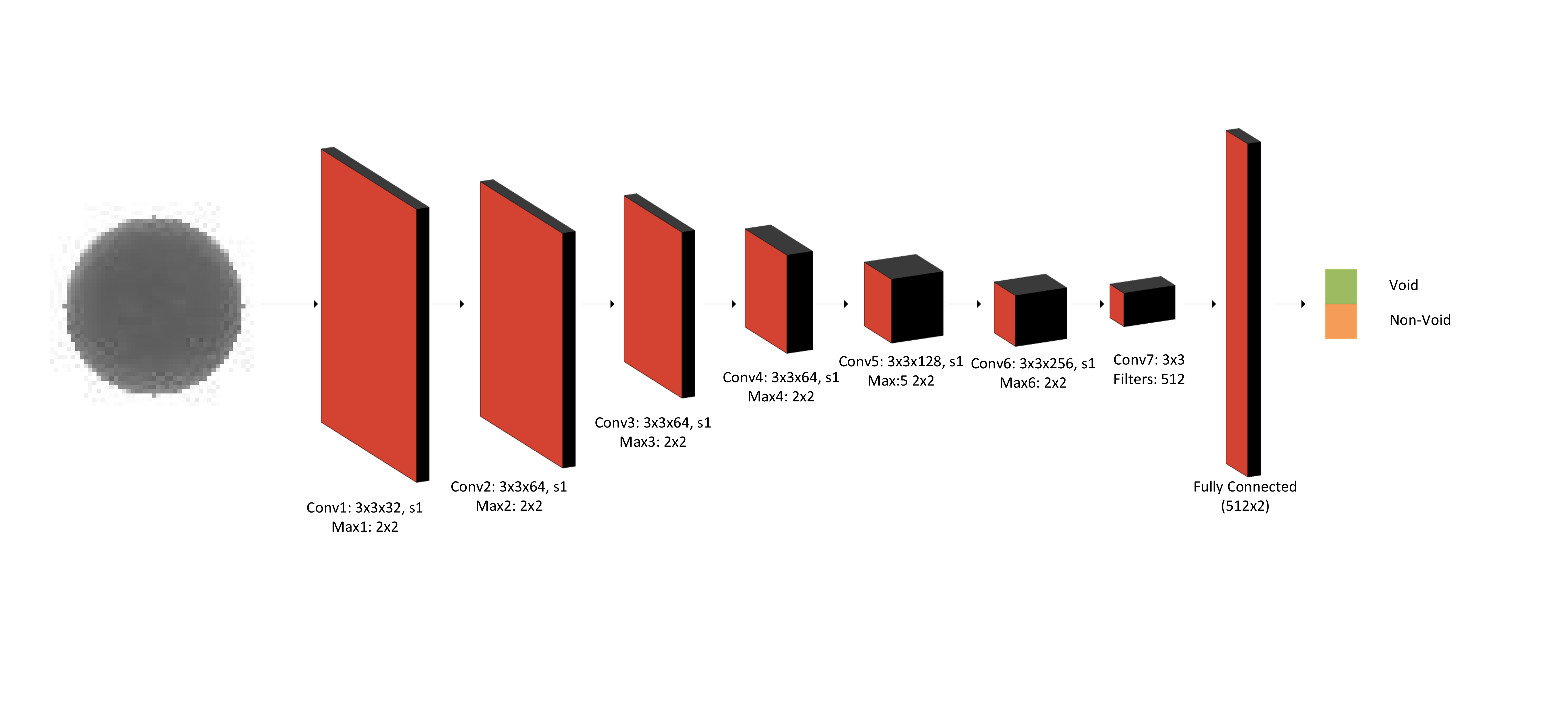}
\caption{Encoder Trained with Fully Connected Layer for Void/Non-Void Binary Classification}
\label{fig:cnn}
\end{figure}

The architecture of U-net is based on encoder and decoder networks. The encoder takes the input image and extracts 
the high level features. The decoder up-sample the extracted features to get the required output. There
are skip between the encoder and decoder that copies the feature map of encoder output in each layer to the 
corresponding layer in decoder. At first, we construct the network for image classification to classify whether
the image consists of void or not. After that, we remove the fully connected layer and add the decoder network to 
construct U-net. Weights for decoder are initialized randomly. The end to end U-net is trained for void segmentation. 
The reason to train the encoder network separately is to have the better initial weights during the training of U-net. 
The classification network and the corresponding network configuration is shown in Fig. \ref{fig:cnn} and 
Tab. \ref{tab:cnn_table}. The input image size is 64x64 and is of gray level. During training, adam optimizer with 
initial learning rate (lr = $10^{-3}$) is used with batch size of 256.

\begin{table}[h!]
  \begin{center}
    \caption{Encoder Network Configuration}
    \label{tab:cnn_table}
    \begin{tabular}{l|c|r} 
      \textbf{Type} & \textbf{Filter size/stride(s)} & \textbf{Output size}\\
      \hline
      Conv1 & 3x3/1 & 64x64x32 \\
      Pool1 & 2x2/1 & 32x32x32 \\
      Conv2 & 3x3/1 & 32x32x64\\
      Pool2 & 2x2/1 & 16x16x64 \\
      Conv3 & 3x3/1 & 16x16x64 \\
      Pool3 & 2x2/1 & 8x8x64 \\
      Conv4 & 3x3/1 & 8x8x64\\
      Pool4 & 2x2/1 & 4x4x64 \\
      Conv5 & 3x3/1 & 4x4x128 \\
      Pool5 & 2x2/1 & 2x2x128 \\
      Conv6 & 3x3/1 & 2x2x256\\
      Pool6 & 2x2/1 & 1x1x256 \\
      Conv7 & 3x3/1 & 1x1x512\\
    \end{tabular}
  \end{center}
\end{table}

\subsubsection{Training U-net}
\begin{figure}[h]
\includegraphics[width=1.0\textwidth, height = 0.8\textheight]{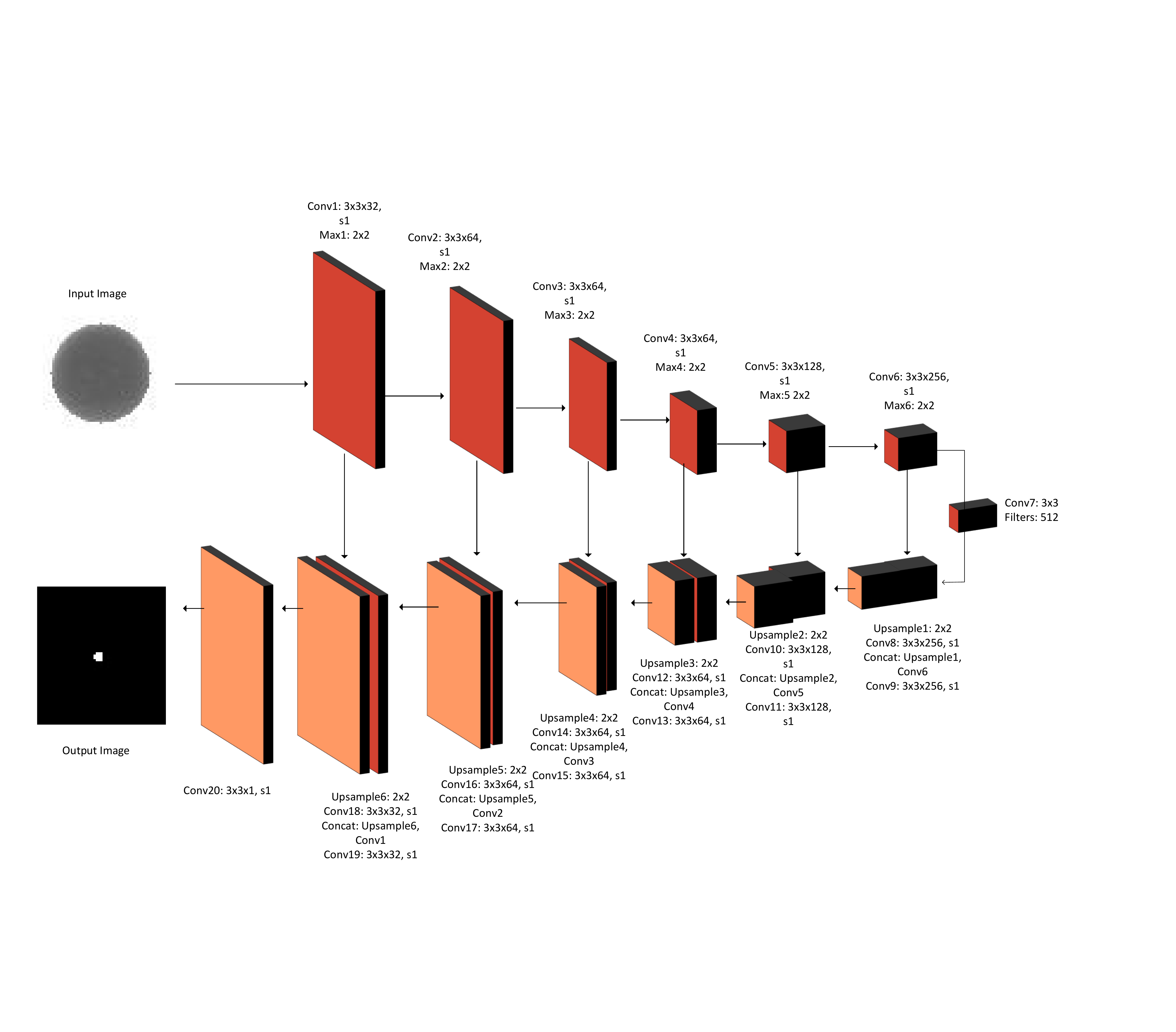}
\caption{End to End Encoder-Decoder Network for Void Segmentation}
\label{fig:unet}
\end{figure}

After training image classification network, full connected layer is removed and the decoder network is added to the
remaining network. The skip connections are added between encoder and decoder layers.    
The U-net architecture is shown in Fig. ~\ref{fig:unet} and the corresponding configuration is shown in 
Tab. ~\ref{tab:unet_table}. The output mask is of same size as input size 64x64. Binary cross entropy is used as 
loss function. The value at each location indicates the probability belonging to void. Adam optimizer with learning 
rate (lr = $10^{-3}$) is used with batch size of 256.

\begin{table}[h!]
  \begin{center}
    \caption{Decoder Network Configuration in U-net}
    \label{tab:unet_table}
    \begin{tabular}{l|c|r} 
      \textbf{Type} & \textbf{Filter size/stride(s)} & \textbf{Output size}\\
      \hline
      Upsample1 & 2x2/1 & 2x2x512 \\
      Conv8 & 3x3/1 & 2x2x256 \\
      Concat(Upsample1, conv6) & - & 2x2x512 \\
      Conv9 & 3x3/1 & 2x2x256 \\
      
      Upsample2 & 2x2/1 & 4x4x256 \\
      Conv10 & 3x3/1 & 4x4x128 \\
      Concat(Upsample2, conv5) & - & 4x4x256 \\
      Conv11 & 3x3/1 & 4x4x128 \\
      
      Upsample3 & 2x2/1 & 8x8x128 \\
      Conv12 & 3x3/1 & 8x8x64 \\
      Concat(Upsample3, conv4) & - & 8x8x128 \\
      Conv13 & 3x3/1 & 8x8x64 \\
      
      Upsample4 & 2x2/1 & 16x16x64 \\
      Conv14 & 3x3/1 & 16x16x64 \\
      Concat(Upsample4, conv3) & - & 16x16x128 \\
      Conv15 & 3x3/1 & 16x16x64 \\
      
      Upsample5 & 2x2/1 & 32x32x64 \\
      Conv16 & 3x3/1 & 32x32x64 \\
      Concat(Upsample5, conv2) & - & 32x32x128 \\
      Conv17 & 3x3/1 & 32x32x64 \\
      
      Upsample6 & 2x2/1 & 64x64x64 \\
      Conv18 & 3x3/1 & 64x64x32 \\
      Concat(Upsample6, conv1) & - & 64x64x64 \\
      Conv19 & 3x3/1 & 64x64x32 \\
      
      Conv20 & 3x3/1 & 64x64x1 \\
      
    \end{tabular}
  \end{center}
\end{table}

\subsection{Testing}

For each segmented soldering ball in test set, the trained U-net is applied to get the probability map that indicates 
the presence of void at each location. The probability map is thresholded with threshold (0.5) to get the binary map. 
The value of 1 indicates the presence of void and for 0 it is the background.   

\subsubsection{Filtering and Void Area Calculation}
After getting the void mask, connected component labelling is applied to separate the individual voids. Regions 
whose area is less than minimum area ($A_{min}$)are filtered out to eliminate the false predictions. The percentage 
of void area with respect to soldering ball is 
calculated as Percentage of Void (\%) = (Area of void/Area of soldering ball) *100. Where, area of void is calculated 
by summing up number of pixels predicted as void. The area of soldering ball is calculated 
as the number of pixels inside the soldering ball.

\section{Experimental Results}
\begin{figure}[H]
\includegraphics[width=1.0\textwidth, height = 0.9\textheight]{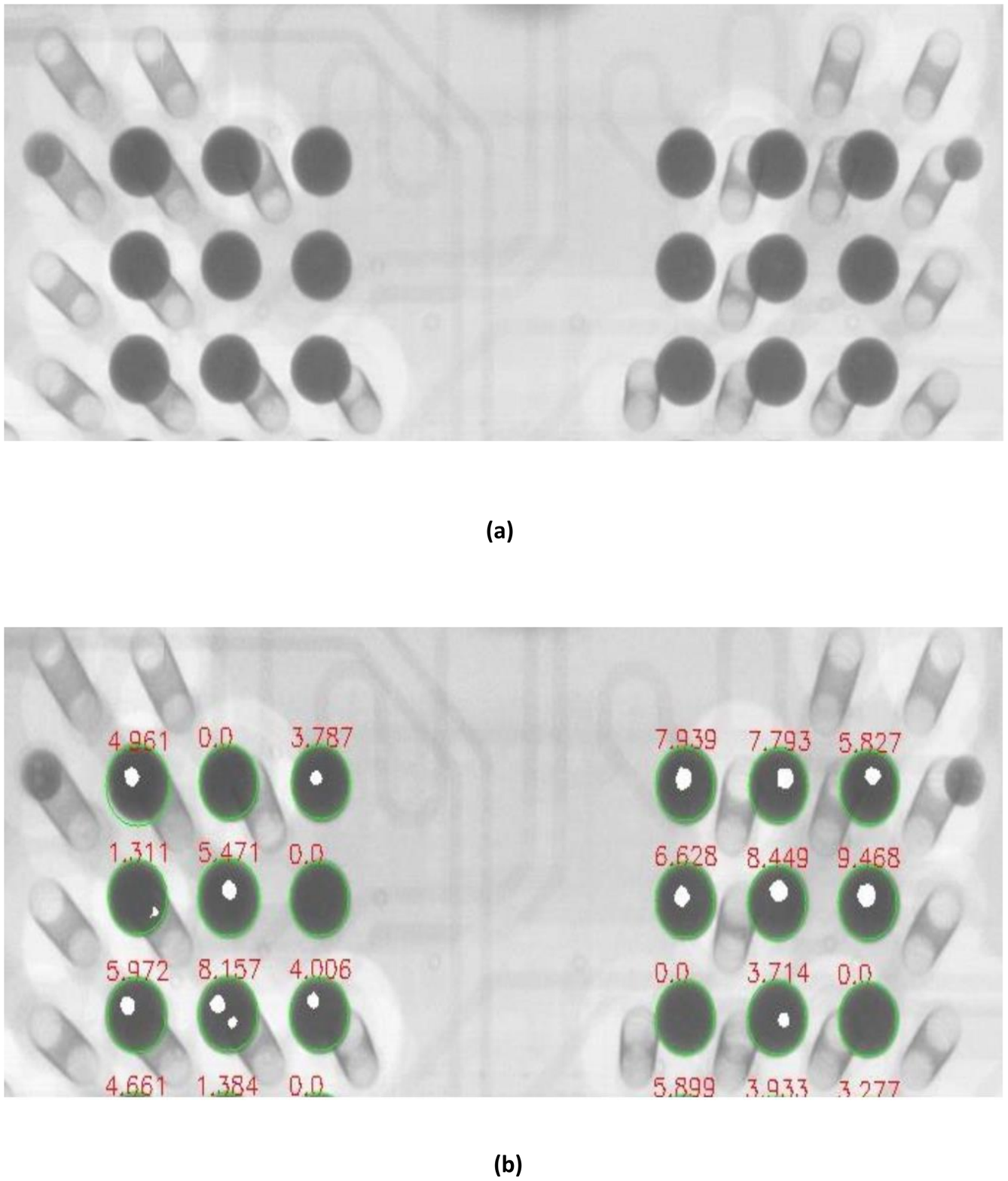}
\caption{(a) Sample Test Image in the Dataset. (b) Detected Voids Overlaid in the Corresponding Test Image}
\label{fig:dataset_sample_and_void}
\end{figure}

In this section we present the experimental results showing the capability of proposed synthetic data and 
void segmentation approach. 
The sample image of the dataset is shown in Fig. \ref{fig:dataset_sample_and_void}(a). The manually annotated and the 
proposed synthetic voids datasets are used to validate the proposed approach. The image resolution of the dataset is  
64x64 where as the diameter of soldering ball is about 40x40. The search range (SR) used
for soldering ball extraction is 5 pixels. $Thr_{min}$ used in separation of void and non-void soldering balls is 6. 
The parameters related to the 
synthetic data generation are $VC_{min}$ = 1 and $VC_{max} = 4$, $VR_{min}$ = 2, $VR_{max}$ = 7, 
$VI_{min}$ = 6, $VI_{max}$ = 9, $VB_{min}$ = 2,$VB_{min}$ = 3, $VN_{min}$ = 1 $VN_{max}$ = 2. The minimum area 
to be considered as void is $A_{min}$ = 9. 

To show the effectiveness of synthetic voids, the comparison results are classified according to the dataset used
for training and testing. The following are the four comparisons used 

1. Train on real voids data and test on real voids data ($Train\_Real\_Test\_Real$) \newline
2. Train on synthetic voids data and test on real voids data ($Train\_Syn\_Test\_Real$) \newline
3. Train on real and synthetic voids data and test on real voids data($Train\_Real\_Syn\_Test\_Real$) \newline

The number of soldering balls in the real dataset consists of 3574. For $Train\_Real\_Test\_Real$,  
the manually annotated data is used for training and testing. For $Train\_Syn\_Test\_Real$, the same synthetic data 
is used for training 
and tested on real data. In $Train\_Real\_Syn\_Test\_Real$, real and synthetic data is used for training 
and tested on real data.  
The precision, recall and F1 score for all the cases is shown in Tab. \ref{tab:pre_rec_four_cases}. 
The detected voids for the sample full image is shown in the Fig. \ref{fig:dataset_sample_and_void}(b). The 
number on top of each soldering ball indicates the percentage of void with respect to the area of soldering ball.    

\begin{table}[h!]
  \begin{center}
    \caption{Precision, Recall and F1 Score Comparison}
    \label{tab:pre_rec_four_cases}
    \begin{tabular}{l|c|c|r} 
      \textbf{} & \textbf{Precision} & \textbf{Recall} & \textbf{F1 score} \\
      \hline
      
      $Train\_Real\_Test\_Real$ & 0.92 & 0.73 & 0.82\\
	  $Train\_Syn\_Test\_Real$ & 0.84 & 0.77 & 0.80 \\
	  $Train\_Real\_Syn\_Test\_Real$ & 0.95 & 0.76 & 0.84 \\
     
    \end{tabular}
  \end{center}
\end{table}

In first case ($Train\_Real\_Test\_Real$), precision is more (false positives are less) and recall 
(false negatives are more). Due to less data, most of the voids are missed in detection resulting in more false
negatives. In second case ($Train\_Syn\_Test\_Real$), the number of false positives are increased due to the domain 
differences between real and synthetic data, irregular voids in the real data. The number of missed detections are 
reduced due to use of synthetic data that covers most of the variations in the dataset. In third case 
($Train\_Real\_Syn\_Test\_Real$), the number of false detections are reduced further, whereas the missed detections
remained same. The reasonable F1 score can be obtained by using only proposed synthetic data compared to the one using 
real data. The network weights trained on synthetic data can be used as initial weights for training on real data with 
less number of real samples. With the help of synthetic data, less number of real samples are needed to get the better 
F1 score.                  

\section{Conclusion}
The quality inspection of BGA by measuring the voids is vital for board yield issues. Various traditional approaches 
were proposed for void segmentation but lacks in scalability and provides less accuracy. In this paper, we applied 
U-net for void segmentation. In addition to that, we proposed an approach to generate the synthetic void dataset 
by considering the variations in real voids. The synthetic data improved the accuracy compared to the ones which
is trained only using real data. The advantage of synthetic data is the data scalability and less number of real
void samples are needed. The proposed system is scalable to various device types or products. There is a scope for
further improvements in the future. In addition to the synthetic voids, synthetic soldering balls can be generated. 
In this paper, we assumed the void shape as circular, but in real the voids can be of any shape. Irregular shape voids
can be generated and augmented for training. Improving the manual annotation can correctly evaluate the importance
of synthetic data. 
\bibliographystyle{unsrtnat}
\bibliography{references}

\end{document}